\documentstyle[11pt]{article}

\global\arraycolsep=1pt 
\input{tcilatex}
\begin{document}

\font\cmss=cmss10 \font\cmsss=cmss10 at 7pt \hfill \hfill CERN-TH/99-183

\hfill 

\vspace{10pt}

\begin{center}
{\large {\bf \vspace{10pt} HIGHER-SPIN\ CURRENT MULTIPLETS \\[0pt]
IN OPERATOR-PRODUCT EXPANSIONS}}

\vspace{10pt}

\bigskip \bigskip 

{\sl Damiano Anselmi}

{\it CERN, Theory Group, CH-1211, Geneva 23, Switzerland}
\end{center}

\vskip 2truecm

\begin{center}
{\bf Abstract}
\end{center}

\vspace{4pt}
Irreducible 
currents with arbitrary spin are
constructed in general space-time dimension,
in the free-field limit and, at the bare level, in the presence
of interactions. 
For the $n$-dimensional generalization of the (conformal) vector
field, the $(n/2-1)$-form is used.
Two-point functions and higher-spin central charges
are evaluated at one loop. 
As an application, 
the higher-spin hierarchies generated by the stress tensor
operator-product expansion are computed
in supersymmetric theories. The results exhibit an
interesting universality.

\vskip 8truecm

CERN-TH/99-183

\vfill\eject 

\setcounter{equation}{0}

\section{Introduction}

The role of higher-spin symmetries in quantum field theory is not
completely clear. On the one side stands the theory of higher-spin fields \cite{hsf}, which consitutes a subject in itself. Recent developments \cite{vasiliev} have shown that a consistent formulation of higher-spin interacting fields can be achieved at the level of the field equations in Anti-de-Sitter space.
On the other side stands the higher-spin symmetry of dynamical fields with ordinary spin (0, 1/2 and 1). It is well known that such a symmetry must be infinite to achieve the algebra closure.
A famous no-go theorem, by Coleman and Mandula \cite{mandula}, asserts
that the S-matrix is trivial in the presence of a mass gap and
a higher-spin symmetry. Then the
theory is free. There remains to understand whether a higher-spin symmetry
does exist if the theory, or just its infrared limit, is conformal and interacting. In this case the Coleman-Mandula theorem does not apply, because the S-matrix is not well defined. Generalizations of the Coleman-Mandula theorem \cite{sohnius}
assert that it is not
possible to extend the group of space-time symmetries beyond the conformal
group in non-trivial theories, but the issue is subtle in the absence of a mass gap.

Higher-spin currents are naturally generated by operator-product expansions
(OPE)
\cite{muta,gatto}. Usually, the currents obtained this way
are not easily classifiable. In supersymmetric theories
it is possible to organize them into higher-spin current multiplets
and hierarchies \cite{n=4,n=2}.
Moreover, powerful theorems \cite{grillo,nach} give rigorous information
about anomalous dimensions, which parametrize the violation of the higher-spin symmetry.

So far, a closed and complete construction of 
irreducible higher-spin currents of spin-0,-1/2 and -1 fields
has not been elaborated. In refs. \cite{berends} the currents
are not irreducible, in \cite{high} irreducible currents are given
for the lowest spins (up to 5 included).

This paper is devoted to the general construction of irreducible 
higher-spin currents and the
computation of their two-point functions. The results are applied
to supersymmetric theories and the hierarchies of current
multiplets
generated by the OPE of the stress tensor are worked-out. The spin
and the space-time dimension are arbitrary, although we 
occasionally focus on
four dimensions for concreteness.

Higher-spin tensor currents \cite{berends} are unique in the free-field
limit if they satisfy the conservation condition and are completely
traceless \cite{high}. In particular, the two-point functions define
the so-called
higher-spin central charges, while the 
higher-spin hierarchies are obtained by diagonalizing the current multiplets
via a combination of supersymmetry and
orthogonalization of the two-point functions \cite{high,n=4,n=2}.
The result is
that the higher-spin current multiplet is {\it universal}, 
the lowest-spin current being 
\[
{\frac{8s(s-1)}{(s+1)(s+2)}}{\cal J}_{s}^{V}+{\frac{2s}{s+1}}{\cal J}%
_{s}^{F}+{\cal J}_{s}^{S},
\]
no matter if the theory is N=4, N=2 or N=1 supersymmetric. Here ${\cal J}%
_{s}^{V,F,S}$ are spin-$s$ currents of vector, fermion and scalar fields, in
the normalization fixed in ref. \cite{high}. 

While the highest-spin current of each multiplet is certainly universal ($8%
{\cal J}_{s}^{V}-2{\cal J}_{s}^{F}+{\cal J}_{s}^{S}$), because it has the
same structure as the stress tensor, the universal character of the
lowest-spin current is less obvious, but still an intrinsic property of
supersymmetry.

This observation raises the interesting question 
about the preservation of such
universality, in a form to be uncovered, when the interaction is
turned on. We recall that the Ferrara--Gatto--Grillo--Nachtmann theorem \cite{grillo,nach} is a
sort of universality theorem for the spectrum of the anomalous dimensions of
the higher-spin currents \cite{n=4,n=2}. For a brief review of these issues, see also \cite{proceeding}.

The currents of this paper, worked out explicitly
in the free-field limit, are easily
extended to interacting theories, at the classical level and, which
is the same, at the bare quantum level. 
This extension is done by
replacing ordinary derivatives by covariant derivatives and extracting all
traces out again. 
The resulting currents are non-conserved, both at the classical
and quantum levels. In families of conformal field theories the violation of
conservation is due to the anomalous dimension. 

We find that various formulas for
fermion and vector fields can be obtained from those of the scalar fields by the
simple replacements $s\rightarrow s-1$, $n\rightarrow n+2$ and $s\rightarrow
s-2$, $n\rightarrow n+4$, respectively, where $s$ is the spin and $n$ is the
space-time dimension. This reduces the computational effort by a factor 3
and justifies working in general dimension.

As the $n$-dimensional generalization of the (conformal) vector
field, we take the $(n/2-1)$-form, whose
field strength admits a decomposition into self-dual
and anti-self-dual component, useful to write down 
conformal currents of arbitrary spin. 

In separate sections we treat: the scalar, fermion and vector fields, 
in sections 2, 3 and 5, respectively, where
the higher-spin currents and their two-point functions are evaluated; 
the N=4 and N=2 hierarchies in
sections 4, 6 and 7, where the mentioned
universality is discussed and it is shown 
that it makes the N=1 hierarchy straightforward; 
useful formulas for projectors and conformal
invariants (sect. 8). In section 9 we compute the
contribution of the $(n/2-1)$-form to the first central charge
$c$ in the trace anomaly and
in section 10 we collect
the complete forms of the higher-spin currents.
The general properties of the $(n/2-1)$-form
in $n$ dimensions are discussed
in the appendix.

\section{Scalar field}

We write the relevant terms of the scalar current of even spin $s$ as

\[
{\cal J}^{(s)}=\sum_{k=0}^{s}a_{k}~\partial ^{(k)}\varphi ~\partial
^{(s-k)}\varphi +b_{k}~\delta _{\cdot \cdot }~\partial _{\alpha
}^{(k)}\varphi ~\partial _{\alpha }^{(s-k)}\varphi +``\delta \delta \text{-}%
{\rm terms}"+\cdots 
\]
with $a_{k}=a_{s-k}$, $b_{k}=b_{s-k},$ $a_{0}=\frac{1}{2}$ and $b_{0}=0$. We
omit the uncontracted indices (or replace them by dots). The symbol $\partial ^{(k)}$ denotes a string of $k$
derivatives; $\partial _{\alpha }^{(k)}$ is a string of $k$ derivatives, one
of which has index $\alpha $. Complete symmetrization in the uncontracted indices is
understood. In order to study conservation and tracelessness it is
sufficient to make one and two indices explicit, respectively.

With one index explicit we have 
\[
{\cal J}_{\mu }^{(s)}=\frac{2}{s}\sum_{k=0}^{s}\left\{ ka_{k}~\partial _{\mu
}^{(k)}\varphi ~\partial ^{(s-k)}\varphi +b_{k}~\delta _{\mu \cdot
}~\partial _{\alpha }^{(k)}\varphi ~\partial _{\alpha }^{(s-k)}\varphi
+(k-1)b_{k}~\delta _{\cdot \cdot }~\partial _{\alpha \mu }^{(k)}\varphi
~\partial _{\alpha }^{(s-k)}\varphi +\cdots \right\} 
\]
and with two indices explicit we have 
\begin{eqnarray*}
{\cal J}_{\mu \nu }^{(s)} &=&\frac{2}{s(s-1)}\sum_{k=0}^{s}\left\{
ka_{k}\left[ (k-1)~\partial _{\mu \nu }^{(k)}\varphi ~\partial
^{(s-k)}\varphi +(s-k)~\partial _{\mu }^{(k)}\varphi ~\partial _{\nu
}^{(s-k)}\varphi \right] \right. \\
&&\!\!\!\!\!\!\!\!\!\!\!\!\!\!\!\!\!\!\!\!{\left. +b_{k}\left[ \delta _{\mu
\nu }~\partial _{\alpha }^{(k)}\varphi ~\partial _{\alpha }^{(s-k)}\varphi
+2(k-1)b_{k}~\delta _{\mu \cdot }~\partial _{\alpha \nu }^{(k)}\varphi
~\partial _{\alpha }^{(s-k)}\varphi +2(k-1)b_{k}~\delta _{\nu \cdot
}~\partial _{\alpha \mu }^{(k)}\varphi ~\partial _{\alpha }^{(s-k)}\varphi
\right] \right\} +\cdots ,}
\end{eqnarray*}
Here the dots stand for terms containing at least one
Kronecker tensor with indices different from $\mu$ and $\nu$.

Tracelessness up to ``$\delta $-terms'' relates $a_{k}$ ad $b_{k}$: 
\begin{equation}
b_{k}=-\frac{k(s-k)}{2s+n-4}a_{k}  \label{bk}
\end{equation}
where $n$ is the space-time dimension. 
In deriving (\ref{bk}) it should be remembered that
$a_k$ and $b_k$ are symmetric
under $k\rightarrow s-k$.
Finally, conservation determines the
coefficients $a_{k}$ recursively, by the condition 
\[
ka_{k}+(s-k+1)a_{k-1}+2b_{k}+2b_{k-1}=0, 
\]
which gives, using (\ref{bk}), 
\begin{equation}
a_{k}=-\frac{(s-k+1)(2s-2k+n-2)}{k(2k+n-4)}a_{k-1},  \label{sn}
\end{equation}
solved by 
\begin{equation}
a_{k}=\frac{(-1)^{k}}{2}\frac{\binom{s}{k}\binom{s+n-4}{k+\frac{n}{2}-2}}{%
\binom{s+n-4}{\frac{n}{2}-2}}.  \label{sol}
\end{equation}
We write our formulas in a convenient way for $n$ even, although the results
are valid for $n$ odd also. In four dimensions we get, in particular, 
\[
a_{k}=\frac{(-1)^{k}}{2}\binom{s}{k}^{2}, 
\]
which can be checked in the case of the stress tensor and the spin-4 scalar
current worked out in ref. \cite{high}.

Summarizing, we have 
\begin{equation}
{\cal J}^{(s)}=\frac{1}{2\binom{s+n-4}{\frac{n}{2}-2}}\sum_{k=0}^{s}(-1)^{k}%
\binom{s}{k}\binom{s+n-4}{k+\frac{n}{2}-2}~\left[ \partial ^{(k)}\varphi
~\partial ^{(s-k)}\varphi -\frac{k(s-k)}{n+2s-4}~\delta _{\cdot \cdot
}~\partial _{\alpha }^{(k)}\varphi ~\partial _{\alpha }^{(s-k)}\varphi
\right]  \label{huge}
\end{equation}
plus terms containing at least two Kronecker tensors.

Now we compute the two-point functions. The form of the correlator is
unique, expressible in terms of suitable projectors or conformal tensors
(see section 8 and \cite{high}). The space-time structure, a sum of various
terms constructed with $x_{\mu }$ and $\delta _{\mu \nu }$, contains a term $%
x_{1}\cdots x_{2s}/|x|^{4s+2n-4}$ with no Kronecker tensor. Unicity assures
that it is sufficient to compute the coefficient of such a term in the
two-point functions, which reduces the effort considerably.

The scalar propagator reads 
\[
\langle \varphi (x)~\varphi (0)\rangle =\frac{\Gamma \left( \frac{n}{2}%
\right) }{2(n-2)\pi ^{\frac{n}{2}}}\frac{1}{|x|^{n-2}}.
\]
We have 
\begin{eqnarray*}
\langle \partial ^{p}\varphi ~\partial ^{s-p}\varphi (x)\quad \partial
^{q}\varphi ~\partial ^{s-q}\varphi (0)\rangle  &=&\frac{x_{1}\cdots x_{2s}}{%
|x|^{4s+2n-4}}\frac{2^{2s-4}(2s)!\left[ \left( \frac{n}{2}-2\right) !\right]
^{2}}{\pi ^{n}}\times  \\
&&\left\{ \frac{\binom{p+q+\frac{n}{2}-2}{\frac{n}{2}-2}\binom{2s-p-q+\frac{n%
}{2}-2}{\frac{n}{2}-2}}{\binom{2s}{p+q}}+\frac{\binom{s-p+q+\frac{n}{2}-2}{%
\frac{n}{2}-2}\binom{s+p-q+\frac{n}{2}-2}{\frac{n}{2}-2}}{\binom{2s}{s-p+q}}%
\right\} 
\end{eqnarray*}
plus $\delta $-terms. 
A factor $(-1)^2$ is omitted, since $s$ is even for the moment.
Therefore the two-point function is 
\[
\langle {\cal J}^{(s)}(x)~{\cal J}^{(s)}(0)\rangle =\frac{x_{1}\cdots x_{2s}%
}{|x|^{4s+2n-4}}\frac{1}{\pi ^{n}}2^{2s-5}(2s)!\left[ \left( \frac{n}{2}%
-2\right) !\right] ^{2}~F[s,n],
\]
where $F[s,n]$ is a double sum of an expression in binomial
coefficients. Precisely,
\begin{equation}
F[s,n]=\sum_{p,q=0}^{s}\frac{(-1)^{p+q}\binom{s}{p}\binom{s}{q}}{\binom{2s}{p+q}%
 \binom{s+n-4}{\frac{n}{2}-2}^{2}}\binom{s+n-4}{%
p+\frac{n}{2}-2}\binom{s+n-4}{q+\frac{n}{2}-2}\binom{p+q+\frac{n}{2}-2}{%
\frac{n}{2}-2}\binom{2s-p-q+\frac{n}{2}-2}{\frac{n}{2}-2}.
\label{gorgo}
\end{equation}
The result is
\begin{equation}
F[s,n]=\frac{(2s+n-4)!~s!}{(2s)!(s+n-4)!},  \label{summa}
\end{equation}
which we have checked with a computer up to $s$ and $n$ equal to 50.
The calculation can be split in two steps. First one considers the sum
\[
\sum_{q=0}^{s}(-1)^{q}\frac{\binom{s}{q}\binom{s+n-4}{%
q+\frac{n}{2}-2}}{\binom{2s}{p+q}}\binom{p+q+\frac{n}{2}-2}{%
\frac{n}{2}-2}\binom{2s-p-q+\frac{n}{2}-2}{\frac{n}{2}-2}
=(-1)^p{s!(s+n-4)!\over (2s)!\left[\left({n\over 2}-2\right)!\right]^2}.
\]
This is the most difficult sum that we meet here. 
Observe that under the replacement $p\rightarrow s-p$ the sum picks up a factor $(-1)^s$.
We have used
Mathematica to compute this sum symbolically for $p=0$ (and $p=s$).
The symbolic calculation, equivalent to the use of the available talbes,
is however more difficult for the other values of $p$.
Once more, we have checked the result numerically 
for other values up to $(s,n)=(50,50)$
and all values of $p\leq s$ in this range.
We do not know how to rigorously show 
that the simple $p$-dependence of this sum
is the claimed one. This
would suffice to complete the proof. 
All our general results for the two-point functions rely on the sum above,
while the other sums that we meet can be treated with the known tables or with
a symbolic calculation. A cross-check of (\ref{summa}) will come from
supersymmetry.

The sum can be written more simply as
\[
\sum_{q=0}^s(-1)^q{(p+q+k)!(2s-p-q+k)!\over q! (s-q)! (q+k)!(s-q+k)!}=(-1)^p,
\]
where $k=n/2-2$.

This result reduces the computation of $F[s,n]$ to
a sum of the form
\begin{equation}
\sum_{k=0}^{s}\binom{s}{k}\binom{s+n-4}{k+\frac{n}{2}-2}={(2s+n-4)!\over 
\left[\left(s+{n\over 2}-2\right)!\right]^2}.
\label{boris}
\end{equation} 

Summarizing,
\begin{equation}
\langle {\cal J}^{(s)}(x)~{\cal J}^{(s)}(0)\rangle =\frac{x_{1}\cdots x_{2s}%
}{|x|^{4s+2n-4}}\frac{2^{2s-5}(2s+n-4)!~s!\left[ \left( \frac{n}{2}-2\right)
!\right] ^{2}}{\pi ^{n}(s+n-4)!}+\delta \text{-terms.}  \label{hy}
\end{equation}
Changing the normalization appropriately, we have checked the agreement with the
results of \cite{high} for $s=0,2,4$ in $n=4$. The above formula, derived
for even spin, extends to odd spin also. This will be shown later on, using
supersymmetry.

The two normalizations, of this paper and of ref. \cite{high}, are related as follows. 
Here we have, from formula (%
\ref{huge}):
\[
{\cal J}^{(s)}={\frac{1}{2}}\frac{(2s+n-4)!\left({n\over 2}-2\right)!}{(s+n-4)!\left( s+\frac{n}{2}%
-2\right) !}\,\varphi \partial ^{s}\varphi +{\rm total\,derivatives},
\]
having used (\ref{boris}).
In ref. \cite
{high} we had 
\[
{\cal J}_{\cite{high}}^{(s)}=\varphi \overleftrightarrow{\partial }%
^{s}\varphi +{\rm total\,derivatives}=2^{s}\varphi \partial ^{s}\varphi +%
{\rm total\,derivatives},
\]
and therefore 
\[
{\cal J}^{(s)}={\frac{1}{2^{s+1}}}\frac{(2s+n-4)!\left({n\over 2}-2\right)!}{(s+n-4)!\left( s+\frac{n}{%
2}-2\right) !}{\cal J}_{\cite{high}}^{(s)},
\]
so that
\begin{equation}
\langle {\cal J}_{\cite{high}}^{(s)}(x)~{\cal J}_{\cite{high}%
}^{(s)}(0)\rangle =\frac{x_{1}\cdots x_{2s}}{|x|^{4s+2n-4}}\frac{%
2^{4s-3}s!(s+n-4)!\left[ \left( s+\frac{n}{2}-2\right) !\right] ^{2}}{\pi ^{n}(2s+n-4)!}
\label{ho}
\end{equation}
plus $\delta $-terms.

\section{Fermion}

We now repeat the construction for the free fermion. Here $s$ can be either
even or odd. The current can be written as

\[
{\cal J}^{(s)}=\sum_{k=0}^{s-1}a_{k}~\partial ^{(k)}\bar{\psi}~\gamma
~\partial ^{(s-k-1)}\psi +b_{k}~\delta _{\cdot \cdot }~\partial _{\alpha
}^{(k)}\bar{\psi}~\gamma ~\partial _{\alpha }^{(s-k-1)}\psi +``\delta \delta 
\text{-}{\rm terms}"+\cdots 
\]
with $a_{k}=(-1)^{s+1}a_{s-k-1}$, $b_{k}=(-1)^{s+1}b_{s-k-1},$ $a_{0}=1$ and $b_{0}=0$. With one index explicit we have 
\begin{eqnarray*}
{\cal J}_{\mu }^{(s)} &=&\frac{1}{s}\sum_{k=0}^{s-1}\left\{ a_{k}~\left[
k\partial _{\mu }^{(k)}\bar{\psi}~\gamma ~\partial ^{(s-k-1)}\psi +\partial
^{(k)}\bar{\psi}~\gamma _{\mu }~\partial ^{(s-k-1)}\psi +(s-k-1)\partial
^{(k)}\bar{\psi}~\gamma ~\partial _{\mu }^{(s-k-1)}\psi \right] \right. \\
&&+b_{k}~\left[ 2\delta _{\mu \cdot }~\partial _{\alpha }^{(k)}\bar{\psi}%
~\gamma ~\partial _{\alpha }^{(s-k-1)}\psi +(k-1)\delta _{\cdot \cdot
}~\partial _{\alpha \mu }^{(k)}\bar{\psi}~\gamma ~\partial _{\alpha
}^{(s-k-1)}\psi \right. \\
&&+\left. \left. \delta _{\cdot \cdot }~\partial _{\alpha }^{(k)}\bar{\psi}%
~\gamma _{\mu }~\partial _{\alpha }^{(s-k-1)}\psi +(s-k-1)\delta _{\cdot
\cdot }~\partial _{\alpha }^{(k)}\bar{\psi}~\gamma ~\partial _{\alpha \mu
}^{(s-k-2)}\psi \right] \right\} +\cdots
\end{eqnarray*}
We do not write the expression with two indices explicit, since by now this
should be straightforward.

Tracelessness implies
\[
b_{k}=-~\frac{k(s-k-1)}{(2s+n-4)}a_{k} 
\]
and conservation gives 
\[
a_{k}~k+a_{k-1}(s-k)+2b_{k}+2b_{k-1}~=0. 
\]
We have the recursion relation 
\[
a_{k}=-\frac{(s-k)(2s-2k+n-2)}{k(2k+n-2)}a_{k-1}, 
\]
which is the same as (\ref{sn}) with $s\rightarrow s-1$ and $n\rightarrow
n+2 $, solved by 
\[
a_{k}=(-1)^{k}\frac{\binom{s-1}{k}\binom{s+n-3}{k+\frac{n}{2}-1}}{\binom{%
s+n-3}{\frac{n}{2}-1}}. 
\]
In particular, we have in four dimensions
\[
a_{k}=\frac{(-1)^{k}}{s+1}\binom{s-1}{k}\binom{s+1}{k+1}.
\]

Summarizing, the fermion current reads 
\begin{equation}
{\cal J}^{(s)}=\sum_{k=0}^{s-1}~\frac{(-1)^{k}\binom{s-1}{k}\binom{s+n-3}{k+%
\frac{n}{2}-1}}{\binom{s+n-3}{\frac{n}{2}-1}}\left[ \partial ^{(k)}\bar{\psi}%
~\gamma ~\partial ^{(s-k-1)}\psi -~\frac{k(s-k-1)}{2s+n-4}~\delta _{\cdot
\cdot }~\partial _{\alpha }^{(k)}\bar{\psi}~\gamma ~\partial _{\alpha
}^{(s-k-1)}\psi \right]  \label{clone}
\end{equation}
plus terms with at least two Kronecker tensors. This expression matches
the complete formulas worked out in \cite{high} up to spin 5.

We now proceed with the computation of the two-point function of the spin-$s$
current. The basic ingredient is 
\begin{eqnarray*}
&&\langle \partial ^{(p)}\bar{\psi}~\gamma ~\partial ^{(s-p-1)}\psi
~(x)\quad \partial ^{(q)}\bar{\psi}~\gamma ~\partial ^{(s-q-1)}\psi
~(0)\rangle  \\
&=&\frac{x_{1}\cdots x_{2s}}{|x|^{4s+2n-4}}\frac{(-1)^{s-1}2^{2s+{\frac{n}{2%
}}-3}}{\pi^{n}}\left(s-p+q+{n\over 2}-2\right)!\left(s+p-q+{n\over 2}-2\right)!,
\end{eqnarray*}
the fermion propagator being 
\[
\langle \psi (x)\,\bar{\psi}(0)\rangle ={\frac{\Gamma \left( {\frac{n}{2}}%
\right) }{2\,\pi ^{\frac{n}{2}}}}{\frac{x\!\!\!\slash
}{|x|^{n}}}.
\]
The two-point function is equal to
\[
\frac{x_{1}\cdots x_{2s}}{|x|^{4s+2n-4}}\frac{1}{\pi ^{n}}2^{2s+{\frac{n}{2}}%
-3}\left[ \left( {\frac{n}{2}}-1\right) !\right] ^{2}(2s-2)!~F[s-1,n+2].
\]
The same sum as in the scalar case appears, and we conclude

\begin{equation}
\langle {\cal J}^{(s)}(x)~{\cal J}^{(s)}(0)\rangle =\frac{x_{1}\cdots x_{2s}%
}{|x|^{4s+2n-4}}\frac{2^{2s+{\frac{n}{2}}-3}(2s+n-4)!~(s-1)!\left[ \left( {%
\frac{n}{2}}-1\right) !\right] ^{2}}{\pi ^{n}(s+n-3)!}  \label{bi}
\end{equation}
plus $\delta $-terms. It can be checked that the sign is in agreement with
reflection positivity and for this purpose it is worth to observe that the odd-spin
currents are ``purely imaginary'' in our notation. The above formula agrees
with the results of \cite{high} up to spin 5 included.

Let us relate the two normalizations. Here we have, from (\ref{clone}) and (%
\ref{boris}):
\[
{\cal J}^{(s)}=\frac{(2s+n-4)!\left({n\over 2}-1\right)!}{(s+n-3)!\left( s+\frac{n}{2}-2\right) !}\,%
\bar{\psi}\gamma \partial ^{s-1}\psi +{\rm total\,derivatives},
\]
while in ref. \cite{high} we had 
\[
{\cal J}_{\cite{high}}^{(s)}=\bar{\psi}\gamma \overleftrightarrow{\partial }%
^{s-1}\psi +{\rm total\,derivatives}=2^{s-1}\bar{\psi}\gamma \partial
^{s-1}\psi +{\rm total\,derivatives},
\]
and therefore $~~~~~~~~~~~~$%
\begin{equation}
~\langle {\cal J}_{\cite{high}}^{(s)}(x)~{\cal J}_{\cite{high}%
}^{(s)}(0)\rangle =\frac{x_{1}\cdots x_{2s}}{|x|^{4s+2n-4}}\frac{2^{4s+{%
\frac{n}{2}}-5}(s+n-3)!(s-1)!\left[ \left( s+\frac{n}{2}-2\right) !\right] ^{2}}{\pi ^{n}(2s+n-4)!}.
\label{bo}
\end{equation}

\section{The simplest hierarchy: hypermultiplet}

The currents that we use are in the normalization of \cite{n=2,high}, namely 
\begin{eqnarray*}
{\cal J}_{s}^{F} &=&\bar{\psi}\gamma \overleftrightarrow{\partial }%
^{s-1}\psi +\text{impr.,}\quad \quad \quad \quad {\cal A}%
_{s}^{F}=\bar{\psi}\gamma _{5}\gamma \overleftrightarrow{\partial }%
^{s-1}\psi +{\rm impr}., \\
{\cal J}_{s}^{S} &=&2\bar{A}_{i}\overleftrightarrow{\partial }^{s}A_{i}+{\rm %
impr}.,
\end{eqnarray*}
where $i=1,2$ and ``impr.'' stands for the improvement terms.

The sypersymmetric transformation relates currents of the same multiplet
according to the formulas 
\begin{eqnarray*}
{\cal J}_{2s}^{S} &\rightarrow &-4~{\cal A}_{2s+1}^{F},\qquad \qquad \\
{\cal J}_{2s}^{F} &\rightarrow &-2~{\cal A}_{2s+1}^{F},\qquad \qquad {\cal A}%
_{2s-1}^{F}\rightarrow -2~{\cal J}_{2s}^{F}+~{\cal J}_{2s}^{S}
\end{eqnarray*}

The general current multiplet of the hierarchy is made of the three currents 
$\Omega _{s}=(\Lambda _{s},{\cal A}_{s+1}^{F},$ $\Xi _{s+2}=-2{\cal J}%
_{s+2}^{F}+{\cal J}_{s+2}^{S})$, where 
\begin{equation}
\Lambda _{s}=-{\frac{s+1}{4(2s+1)}}\left( {\frac{2s}{s+1}}{\cal J}_{s}^{F}+%
{\cal J}_{s}^{S}\right)  \label{uno}
\end{equation}
($s$ even), fixed by imposing orthogonality between the currents with the
same spin, $\Lambda _{s}$ and $\Xi _{s}$. The results of \cite{n=2}
are correctly reproduced.

\section{Vector field and its generalization to arbitrary
dimension}

The vector field $A_\mu$ is not conformal in dimension different from 4.
As the
generalization of the 
four-dimensional vector field to arbitrary
(even) dimension,
it is better to take the $(n/2-1)$-form
$A_{\mu_1\ldots \mu_{l-1}}$, where $s=2l$.  
This theory is indeed conformal.
The $l-1$ indices are completely antisymmetrized and this ``vector''
has the usual gauge-invariance. The field strength reads
\[
F_{\mu_1\ldots \mu_l}=\partial_{\mu_1} A_{\mu_2\ldots \mu_l }-
\partial_{\mu_2}A_{\mu_1\mu_3\ldots \mu_l}-\partial_{\mu_3}
A_{\mu_2\mu_1\mu_4\ldots \mu_l}+\cdots
\]
and the lagrangian is
\[
{\cal L}={1\over n}F^2_{\mu_1\ldots \mu_l}\equiv {1\over n}F^2.
\]
The stress tensor,
\[
T_{\mu\nu}=F_{\mu\{\alpha\}}F_{\nu\{\alpha\}}-{1\over n}\delta_{\mu\nu}
F^2,
\]
is traceless, as claimed.
Here $\{\alpha\}$ denotes collectively a string of $l-1$ indices
and repeated $\{\alpha\}$'s denote contracted indices. The contraction
is done with the identity in the space of tensors with
$(l-1)$-antisymmetric indices (see the Appendix).

\vskip .2truecm

We can now proceed to construct the higher-spin currents.
As we explain in the Appendix, 
the four-dimensional formalism extends straightforwardly
to arbitrary dimension multiple of 4, but also
$l$=odd does not present problems
if complex fields are considered. We therefore assume that $l$=even
and the results will extend immediately to $l$=odd.
In particular, the stress tensor can be written as
\[
T_{\mu\nu}=2F^+_{\mu\{\alpha\}}F^-_{\nu\{\alpha\}},
\]
without care about the positions of the indices $\mu$ and $\nu$.
The same can be said of the currents with arbitrary spin.
Conservation and tracelessness are evident.

We write the relevant terms of the current of spin $s$ as
\[
{\cal J}^{(s)}=\sum_{k=0}^{s-2}a_{k}~\partial ^{(k)}F_{\{\alpha\} }^{+}~\partial
^{(s-k-2)}F_{\{\alpha\} }^{-}+b_{k}~\delta _{\cdot \cdot }~\partial _{\beta
}^{(k)}F_{\{\alpha\} }^{+}~\partial _{\beta }^{(s-k-2)}F_{\{\alpha\} }^{-}+``\delta
\delta \text{-}{\rm terms}"+\cdots, 
\]
with $a_{k}=(-1)^sa_{s-k}$, $b_{k}=(-1)^{s}b_{s-k},$ $a_{0}=1$ and $b_{0}=0$. $%
F_{\{\alpha\} }^{\pm }$ are the self-dual and anti-self-dual components of the
field strength, where one index is not written explicitly
(for example, $F^+_{\{\alpha\}} F^-_{\{\alpha\}}$ 
stands for $F^+_{\{\alpha\}\mu}F^-_{\{\alpha\}\nu}$
making the two indices, $\mu$ and $\nu$, explicit). 
Here $\{\alpha\}$ denotes a string of (contracted) $l-1$ indices,
as before.

With one index explicit we have 
\begin{eqnarray*}
{\cal J}_{\mu }^{(s)} &=&\frac{1}{s}\sum_{k=0}^{s-2}\left\{ a_{k}\left[
k\partial _{\mu }^{(k)}F_{\{\alpha\} }^{+}~\partial ^{(s-k-2)}F_{\{\alpha\}
}^{-}\right.\right.\\
&&\left.+ (s-k-2)\partial ^{(k)}F_{\{\alpha\} }^{+}~\partial _{\mu
}^{(s-k-2)}F_{\{\alpha\} }^{-}+2\partial ^{(k)}F_{\{\alpha\} \mu }^{+}~\partial ^{(s-k-2)}F_{\{\alpha\}
}^{-}\right] \\
&&+b_{k}~\left[ 2\delta _{\mu \cdot }~\partial _{\beta }^{(k)}F_{\{\alpha\}
}^{+}~\partial _{\beta }^{(s-k-2)}F_{\{\alpha\} }^{-}+\right. 2\delta _{\cdot
\cdot }~\partial _{\beta }^{(k)}F_{\{\alpha\} \mu }^{+}~\partial _{\beta
}^{(s-k-2)}F_{\{\alpha\} }^{-}\\
&&+(k-1)\delta _{\cdot \cdot }~\partial _{\beta \mu }^{(k)}F_{\{\alpha\}
}^{+}~\partial _{\beta }^{(s-k-2)}F_{\{\alpha\} }^{-}+\left. \left.
(s-k-3)\delta _{\cdot \cdot }~\partial _{\beta }^{(k)}F_{\{\alpha\}
}^{+}~\partial _{\beta \mu }^{(s-k-2)}F_{\{\alpha\} }^{-}\right] \right\} .
\end{eqnarray*}
One obtains the same relations as in the scalar case, with $s\rightarrow s-2$
and $n\rightarrow n+4$. Therefore 
\begin{equation}
a_{k}=(-1)^{k}\frac{\binom{s-2}{k}\binom{s+n-2}{k+\frac{n}{2}}}{\binom{s+n-2%
}{\frac{n}{2}}},\qquad b_{k}=-\frac{k(s-k-2)}{2s+n-4}a_{k}.
\end{equation}
In four dimensions, in particular, 
\[
a_{k}=\frac{2(-1)^{k}}{(s+1)(s+2)}\binom{s-2}{k}\binom{s+2}{k+2}, 
\]
which can be checked in the cases worked out explicitly in ref. \cite{high}.

Summarizing, we have
\[
{\cal J}^{(s)}=\sum_{k=0}^{s-2}\frac{(-1)^{k}\binom{s-2}{k}
\binom{s+n-2}{k+{n\over 2}}%
}{\binom{s+n-2}{{n\over 2}}}~\left[ \partial ^{(k)}F_{\{\alpha\} }^{+}~\partial
^{(s-k-2)}F_{\{\alpha\} }^{-}-\frac{k(s-k-2)}{2s+n-4}~\delta _{\cdot \cdot
}~\partial _{\beta }^{(k)}F_{\{\alpha\} }^{+}~\partial _{\beta
}^{(s-k-2)}F_{\{\alpha\} }^{-}\right] 
\]
plus terms containing at least two Kronecker tensors.

For the calculation of the two-point functions we need the result 
\begin{eqnarray*}
\langle \partial ^{(p)}F_{\{\alpha\} }^{+}\partial ^{(s-p-2)}F_{\{\alpha\}
}^{-}(x)\,\,\partial ^{(q)}F_{\{\beta\} }^{+}\partial ^{(s-q-2)}F_{\{\beta\}
}^{-}\rangle =~~~~~~~~~~~~~~~~~~~~~~~~~~~~~~~~~\\
{\frac{x_1\cdots x_{2s}}{|x|^{4s+2n-4}}}
{(-1)^s2^{2s-7}n^2(n-2)!(2s-4)!\over \pi^n}
{\binom{s-p+q+{n\over 2}-2}{{n\over 2}} 
\binom{s+p-q+{n\over 2}-2}{{n\over 2}} 
\over 
\binom{2s-4}{s-p+q-2}},
\end{eqnarray*}
up to $\delta$-terms as usual. The derivation of this formula
is lengthy, but does not present particular difficulties.
See the appendix for the propagator and other details.

The result involves a binomial sum similar to the one of the fermion
and scalar currents. Precisely,
\begin{eqnarray}
\langle {\cal J}^{(s)}(x)\,{\cal J}^{(s)}(0)\rangle &=&{\frac{x_{1}\cdots
x_{2s}}{|x|^{4s+2n-4}}}{2^{2s-7}n^2(n-2)!(2s-4)!\over \pi^n}
F[s-2,n+4]
=\nonumber\\&&
\frac{x_{1}\cdots
x_{2s}}{|x|^{4s+2n-4}}\frac{2^{2s-7}n^2(n-2)!}{\pi^n}
{(2s+n-4)!(s-2)!\over (s+n-2)!}.  
\label{ci}
\end{eqnarray}

We now check the values of \cite{high}. We can write 
\[
{\cal J}^{(s)}={\frac{\left({n\over 2}\right)!
(2s+n-4)!}{(s+n-2)!\left(s+{n\over 2}-2\right)!}}F_{\{\alpha\} }^{+}\partial
^{s-2}F_{\{\alpha\} }^{-}+{\rm total\,derivs.}=
-{\frac{2^{2-s}\left({n\over 2}\right)!
(2s+n-4)!}{(s+n-2)!\left(s+{n\over 2}-2\right)!}}
{\cal J}_{\cite{high}}^{(s)},
\]
where we recall that ${\cal J}_{\cite{high}}^{(s)}=-F_{\{\alpha\} }^{+}%
\overleftrightarrow{\partial }^{s-2}F_{\{\alpha\} }^{-}$ plus improvement terms.
The two-point function in the notation of \cite{high} reads 
\begin{equation}
\langle {\cal J}_{\cite{high}}^{(s)}(x)\,{\cal J}_{\cite{high}%
}^{(s)}(0)\rangle ={\frac{x_{1}\cdots x_{2s}}{|x|^{4s+4}}}
{\frac{
2^{4s-9}(n-2)!}{\pi^n\left[\left({n\over 2}-1\right)!\right]^2}}
{\frac{(s-2)!(s+n-2)!\left[\left(s+{n\over 2}-2\right)!\right]^2}
{(2s+n-4)!}}.
\label{co}
\end{equation}

\section{The N=4 hierarchy}

Now we construct the complete N=4 hierarchy, according to the results of 
\cite{n=4}. We normalize the currents as in \cite{high,n=4}:
\begin{eqnarray}
{\cal J}_{s}^{V} &=&-F_{\alpha }^{+}\overleftrightarrow{\partial }%
^{s-2}F_{\alpha }^{-}+{\rm impr}.,\quad \quad {\cal J}_{s}^{F}=\frac{1}{2}%
\bar{\lambda}_{i}\gamma \overleftrightarrow{\partial }^{s-1}\lambda _{i}+%
{\rm impr}.,  \label{n4} \\
{\cal A}_{s}^{V} &=&-F_{\alpha }^{+}\overleftrightarrow{\partial }%
^{s-2}F_{\alpha }^{-}+{\rm impr}.,\quad \quad {\cal A}_{s}^{F}=\frac{1}{2}%
\bar{\lambda}_{i}\gamma _{5}\gamma \overleftrightarrow{\partial }%
^{s-1}\lambda _{i}+{\rm impr}.,  \nonumber \\
{\cal J}_{s}^{S} &=&A_{ij}\overleftrightarrow{\partial }^{s}A_{ij}+B_{ij}%
\overleftrightarrow{\partial }^{s}B_{ij}+{\rm impr}.,  \nonumber
\end{eqnarray}
where $i=1,\ldots 4$, ${\cal J}$ and ${\cal A}$ are even-spin and odd-spin
currents, respectively. The supersymmetry operation is 
\begin{eqnarray*}
{\cal A}_{2s-1}^{V} &\rightarrow &-4~{\cal J}_{2s}^{V}+\frac{1}{4}~{\cal J}%
_{2s}^{F},\qquad \qquad {\cal A}_{2s-1}^{F}\rightarrow -16~{\cal J}%
_{2s}^{V}-2~{\cal J}_{2s}^{F}+2~{\cal J}_{2s}^{S}, \\
{\cal J}_{2s}^{V} &\rightarrow &-4~{\cal A}_{2s+1}^{V}+\frac{1}{4}~{\cal A}%
_{2s+1}^{F},\qquad \qquad {\cal J}_{2s}^{F}\rightarrow -16~{\cal A}%
_{2s+1}^{V}-2~{\cal A}_{2s+1}^{F}, \\
{\cal J}_{2s}^{S} &\rightarrow &-6~{\cal A}_{2s+1}^{F}.
\end{eqnarray*}

The general structure of the multiplet is 
\begin{eqnarray*}
\Sigma _{s} &=&a_{s}~{\cal J}_{s}^{V}+b_{s}~{\cal J}_{s}^{F}+c_{s}~{\cal J}%
_{s}^{S} \\
\Lambda _{s+1} &=&-4(a_{s}+4b_{s})~{\cal A}_{s+1}^{V}+{\frac{1}{4}}%
(a_{s}-8b_{s}-24c_{s})~{\cal A}_{s+1}^{F} \\
\Xi _{s+2} &=&12(a_{s}+8b_{s}+8c_{s})~{\cal J}_{s+2}^{V}-{\frac{3}{2}}%
(a_{s}-8c_{s})~{\cal J}_{s+2}^{F}+{\frac{1}{2}}(a_{s}-8b_{s}-24c_{s})~{\cal J%
}_{s+2}^{S} \\
\Omega _{s+3} &=&3(a_{s}+16b_{s}+24c_{s})\left( -8~{\cal A}_{s+3}^{V}+{\cal A%
}_{s+3}^{F}\right) \\
\Upsilon _{s+4} &=&6(a_{s}+16b_{s}+24c_{s})\left( 8~{\cal J}_{s+4}^{V}-2~%
{\cal J}_{s+4}^{F}+{\cal J}_{s+4}^{S}\right) ,
\end{eqnarray*}
where $s$ is any even integer number. In addition, the
stress tensor  is a current multiplet in itself and is in practice $
\Upsilon _{2}$. Note that the capital Greek letters are not used here in the same
way as in ref. \cite{n=4}.

We conventionally fix the normalization so that the scalar component of the
highest-spin current has coefficient 1: 
\[
6(a_s+16 b_s+24 c_s)=1. 
\]

Secondly, we impose that multiplets with different spins be orthogonal. The
conditions $\langle \Upsilon_s\, \Sigma_s\rangle =0$ and $\langle
\Sigma_{s+2}\, \Xi_{s+2}\rangle =0$ give, respectively, 
\begin{eqnarray*}
0&=&{\frac{(s+1)(s+2)}{4s(s-1)}}a_s-4{\frac{s+1}{s}}b_s+6c_s, \\
0&=&{\frac{(s+3)(s+4)}{8(s+1)(s+2)}}(a_s+8b_s+8c_s)a_{s+2}- {\frac{s+3}{s+2}}%
(a_s-8c_s)b_{s+2}+(a_s-8b_s-24c_s)c_{s+2}.
\end{eqnarray*}

The three conditions are solved uniquely by $a_{s},b_{s},c_{s}$ such that 
\begin{equation}
\Sigma _{s}={\frac{(s+1)(s+2)}{96(2s+1)(2s+3)}}\left[ {\frac{8s(s-1)}{%
(s+1)(s+2)}}{\cal J}_{s}^{V}+{\frac{2s}{s+1}}{\cal J}_{s}^{F}+{\cal J}%
_{s}^{S}\right] ,  \label{due}
\end{equation}
in agreement with the results of \cite{n=4}. The remaining orthogonality
relationships are automatically satisfied, which is also a
cross-check of the formula for $F[s,n]$.

The orthogonalized two-point functions define the so-called ``higher-spin
central charges'', in particular 
\begin{equation}
\langle \Sigma _{s}(x)\,\Sigma _{s}(0)\rangle ={\frac{%
2^{4s-10}(s!)^2(s+1)!(s+2)!}{9(2s+3)!}} \left( {\frac{x_{1}\cdots x_{2s}%
}{|x|^{4s+4}}}+\delta \text{-terms}\right) \text{.}  \label{rima}
\end{equation}
Note that the curious $s\rightarrow \infty $ limit 
\[
\Sigma _{\infty }={\frac{1}{384}}\left( {8}{\cal J}_{s}^{V}+{2}{\cal J}%
_{s}^{F}+{\cal J}_{s}^{S}\right) . 
\]

\section{The N=2 vector multiplet}

We conclude the applications with the hierarchy of the N=2
vector multiplet \cite{n=2}. The fermion and vector currents are the same as
in (\ref{n4}) (with $i=1,2$ now) and the scalar currents are

\begin{equation}
{\cal J}_{s}^{S}=M\overleftrightarrow{\partial }^{s}M+N\overleftrightarrow{%
\partial }^{s}N+{\rm impr}.,\qquad \quad {\cal A}_{s}^{S}=-2iM\overleftrightarrow{%
\partial }^{s}N+{\rm impr}..  \label{norma}
\end{equation}
We also have odd-spin currents for scalar fields. We determine their
two-point functions using supersymmetry, as in \cite{n=2}.

The supersymmetry operation is 
\begin{eqnarray*}
{\cal J}_{2s}^{S} &\rightarrow &-2~{\cal A}_{2s+1}^{F}+2~{\cal A}%
_{2s+1}^{S},\qquad \qquad {\cal A}_{2s-1}^{S}\rightarrow -2~{\cal J}%
_{2s}^{F}+2~{\cal J}_{2s}^{S}, \\
{\cal J}_{2s}^{F} &\rightarrow &-8~{\cal A}_{2s+1}^{V}+~{\cal A}%
_{2s+1}^{S},\qquad \qquad {\cal A}_{2s-1}^{F}\rightarrow -8~{\cal J}%
_{2s}^{V}+~{\cal J}_{2s}^{S}, \\
{\cal J}_{2s}^{V} &\rightarrow &-2~{\cal A}_{2s+1}^{V}+\frac{1}{4}~{\cal A}%
_{2s+1}^{F},\qquad \qquad {\cal A}_{2s-1}^{V}\rightarrow -2~{\cal J}%
_{2s}^{V}+\frac{1}{4}~{\cal J}_{2s}^{F}.
\end{eqnarray*}
Writing 
\begin{eqnarray*}
\Sigma _{s} &=&~{a_{s}~{\cal J}_{s}^{V}+b_{s}~{\cal J}_{s}^{F}+c_{s}~{\cal J}%
_{s}^{S},}\qquad \\
\Lambda _{s+1} &=&-2(a_{s}+4b_{s})~{\cal A}_{s+1}^{V}+{\frac{1}{4}}%
(a_{s}-8c_{s})~{\cal A}_{s+1}^{F}+(b_{s}+2c_{s})~{\cal A}_{s+1}^{S}, \\
\Xi _{s+2} &=&{\frac{a_{s}+8b_{s}+8c_{s}}{4}}\left( 8{\cal J}_{s+2}^{V}-2~%
{\cal J}_{s+2}^{F}+~{\cal J}_{s+2}^{S}\right) ,
\end{eqnarray*}
(${\cal J}\leftrightarrow {\cal A}$ when $s$ is odd) we impose 
the normalization $%
a_{s}+8b_{s}+8c_{s}=4$ and orthogonality ($\langle \Sigma
_{s+1}~\Lambda _{s+1}\rangle =\langle \Sigma _{s}~\Xi _{s}\rangle =0$). The
result is

\begin{equation}
\Sigma _{s}={\frac{(s+2)}{4(2s+1)}}\left[ {\frac{8s(s-1)}{(s+1)(s+2)}}{\cal J%
}_{s}^{V}+{\frac{2s}{s+1}}{\cal J}_{s}^{F}+{\cal J}_{s}^{S}\right] .
\label{tre}
\end{equation}

Comparing (\ref{uno}), (\ref{due}) and (\ref{tre}) we observe an unexpected
universality of the higher-spin hierarchy. The lowest spin is always 
\begin{equation}
{\frac{8s(s-1)}{(s+1)(s+2)}}{\cal J}_{s}^{V}+{\frac{2s}{s+1}}{\cal J}%
_{s}^{F}+{\cal J}_{s}^{S},  \label{struc}
\end{equation}
and theories with different supersymmetric contents just differ in
restrictions such as the parity of $s$, the presence or absence of the vector currents ${\cal J}%
_{s}^{V}$, the overall normalization, etc.. For example, $s$ takes any integer value
in the N=2 vector multiplet, but only even values in the N=4 multiplet and in the
hypermultiplet.

Observe that the highest-spin component of the current multiplets ($8{\cal J%
}_{s}^{V}-2{\cal J}_{s}^{F}+{\cal J}_{s}^{S}$) is also universal and
independent of the spin. This follows from the universality of the stress
tensor. The highest-spin currents are annihilated, by definition, by the supersymmetry
transformation and the unique structure with this property
is that of the stress tensor.

In conclusion, the infinite OPE algebra of (supersymmetric) conformal field
theory in higher dimensions is uniquely described by the simple, universal,
structure (\ref{struc}). It is plausible that the universality that we have
uncovered is much deeper. It should be an intrinsic property of
supersymmetry, since the highest-spin/lowest-spin two-point function ($%
\langle\Upsilon_s\,\Sigma_s\rangle$ for N=4 and $\langle\Xi_s\,\Sigma_s%
\rangle$ for N=2) is proportional to the difference $2N_V-4N_F+N_S$ between
the number of bosonic and fermionic degrees of freedom ($N_{V,F,S}$ being
the number of vectors, Dirac fermions and real scalars, respectively).

This observation extends immediately to N=1 hierarchies. Indeed, N=1
supersymmetric multiplets contain only two types of fields (vector, fermion
for the vector multiplet and fermion, scalar for the scalar multiplet), and
the universality of the highest-spin current, together with
orthogonality, implies immediately
the universality of the lowest-spin current. This fact is not obvious,
instead, in extended supersymmetry. The current multiplets are
\[
\matrix{{\frac{8s(s-1)}{(s+1)(s+2)}}{\cal J}_{s}^{V}+{\frac{2s}{s+1}}{\cal J}%
_{s}^{F}\cr
8{\cal J}_{s+1}^{V}-2~{\cal J}_{s+1}^{F}},~~~~~~~~~~~~~~~~~~~\matrix{{\frac{%
2s}{s+1}}{\cal J}_{s}^{F}+{\cal J}_{s}^{S}\cr
-2~{\cal J}_{s+1}^{F}+~{\cal J}_{s+1}^{S}},
\]
for the vector and scalar multiplets, respectively,
up to normalization factors. Here $s$ takes any integer value.

A byproduct of the calculations of this section is the extension of the
scalar two-point functions, (\ref{hy}) and (\ref{ho}), which we have derived
for even spin, to odd spin (and complex scalar fields), with the normalization
specified in (\ref{norma}). This result is implied by orthogonality of the
odd-spin components of the N=2 vector-field current multiplets.

\section{Formulas for projectors and conformal tensors}

In this section we work out formulas to express the two-point functions in
the notation of \cite{high} and write them down in complete form 
(including the $\delta $-terms).

We start from the projectors ${\prod }_{\mu _{1}\cdots \mu _{s},\nu
_{1}\cdots \nu _{s}}^{(s)}$, which are defined as the unique $2s$ indiced
polynomials of degree $2s$ in derivatives, completely symmetric, conserved
and traceless in $\mu _{1}\cdots \mu _{s}$ and $\nu _{1}\cdots \nu _{s}$. We
write them in compact notation as ${\prod }_{\{\mu \},\{\nu \}}^{(s)}$ and
we expand them as 
\begin{equation}
{\prod }_{\{\mu \},\{\nu \}}^{(s)}=\sum_{k=0}^{\left[ \frac{s}{2}\right]
}a_{k}\pi _{\bar\mu\bar\mu}^{k}\pi _{\bar\nu\bar\nu}^{k}\pi
_{\bar\mu\bar\nu}^{s-2k}.  \label{espre}
\end{equation}
Here $\pi _{\bar\mu\bar\mu}^{k}$ denotes a string of $k$ projectors $\pi
_{\alpha \beta }=\partial _{\alpha }\partial _{\beta }-\delta _{\alpha \beta
}\Box $, with indices from the set $\{\mu \}$. The normalization is
conventionally fixed so that $a_{0}=1$. Expression (\ref{espre}) is
automatically conserved and we have to impose tracelessness.
For this purpose, we can make two indices of the set $\{\mu \} $, say $%
\alpha $ and $\beta $, explicit: 
\begin{eqnarray*}
{\prod }_{\{\mu \},\{\nu \}}^{(s)} &=&\sum_{k=0}^{\left[ \frac{s}{2}\right]
}a_{k}\pi _{\bar\nu\bar\nu}^{k}\left[ 2k~\pi _{\alpha \beta }\pi
_{\bar\mu\bar\mu}^{k-1}\pi _{\bar\mu\bar \nu }^{s-2k}+2k(s-2k)(\pi _{\alpha
\bar\mu}\pi _{\beta \bar \nu }+\pi _{\alpha \bar\nu}\pi _{\beta \bar\mu})\pi
_{\bar\mu\bar\mu}^{k-1}\pi _{\bar\mu\bar\nu}^{s-2k-1}\right. \\
&&\left. +4k(k-1)\pi _{\alpha \bar\mu}\pi _{\beta \bar\mu}\pi
_{\bar\mu\bar\mu}^{k-2}\pi _{\bar\mu\bar \nu }^{s-2k}+(s-2k)(s-2k-1)\pi
_{\alpha \bar\nu}\pi _{\beta \bar\nu}\pi _{\bar\mu\bar\mu}^{k}\pi _{\bar\mu
\bar\nu}^{s-2k-2}\right] .
\end{eqnarray*}
Tracing in $\alpha $ and $\beta $ we get the recursion relation 
\begin{equation}
a_{k}=-a_{k-1}\frac{(s-2k+2)(s-2k+1)}{2k(2s-2k+n-3)}  \label{n}
\end{equation}
in generic dimension $n$. The solution is 
\[
a_{k}=(-1)^{k}\frac{(2s-2k+n-4)!~s!~\left( s+\frac{n}{2}-2\right) !}{%
k!~(s-2k)!~\left( s-k+\frac{n}{2}-2\right) !~(2s+n-4)!}, 
\]
which correctly reproduces the values of \cite{high} for $n=4$ and $%
s=0,\ldots, 5$.

It can be shown, using the known tables, that 
\[
\sum_{k=0}^{\left[ \frac{s}{2}\right] }a_{k}=\frac{2^{s}(s+n-4)!\left( s+{%
\frac{n}{2}}-2\right) !}{\left( {\frac{n}{2}}-2\right) !(2s+n-4)!}
\]
and the general space-time structure of the two-point function $%
\langle {\cal J}_{\{\mu \}}^{(s)}(x)\,{\cal J}_{\{\nu \}}^{(s)}(0)\rangle $
of the spin-$s$ currents is 
\[
{\prod }_{\{\mu \},\{\nu \}}^{(s)}\left( \frac{1}{|x|^{2n-4}}\right) =\frac{%
x_{1}\cdots x_{2s}}{|x|^{4s+2n-4}}\frac{2^{3s}(2s+n-3)!(s+n-4)!\left( s+{\frac{n%
}{2}}-2\right) !}{(n-3)!\left( {\frac{n}{2}}-2\right) !(2s+n-4)!}+\delta 
\text{-terms.}
\]
This allows us to convert the formulas of the previous sections, (\ref{hy}),
(\ref{ho}), (\ref{bi}), (\ref{bo}), (\ref{ci}) and (\ref{co}), in the more
elegant notation of \cite{high}. For example (\ref{rima}) can be
re-expressed as 
\[
\langle \Sigma _{s}(x)\,\Sigma _{s}(0)\rangle ={\frac{%
2^{s-10}(s+1)!(s+2)!}{9(2s+1)(2s+3)!}}{\prod }^{(s)}\left( 
\frac{1}{|x|^{4}}\right) .
\]

An equivalent way to express the two-point functions is by using the
conformal tensor ${\cal I}_{\mu _{1}\cdots \mu _{s},\nu _{1}\cdots \nu
_{s}}^{(s)}(x)$, first introduced by Ferrara {\it et al}. in refs. \cite{gatto}%
, which we write in compact form as 
\begin{equation}
{\cal I}_{\{\mu \},\{\nu \}}^{(s)}(x)=\sum_{k=0}^{\left[ \frac{s}{2}\right]
}b_{k}\delta _{\bar\mu\bar\mu}^{k}\delta _{\bar\nu\bar\nu}^{k}{\cal I}%
_{\bar\mu\bar\nu}^{s-2k}(x),  \label{boh}
\end{equation}
where the powers $k$ and $s-2k$ denote the number of Kronecker tensors and
the number of tensors ${\cal I}_{\mu \nu }(x)=\delta _{\mu \nu }-2x_{\mu
}x_{\nu }/|x|^{2}$, respectively. The tensor ${\cal I}_{\mu \nu }(x)$ \cite{schreier}
is indeed the building block of all conformal tensors. We normalize (\ref{boh})
with $b_{0}=1$. The formal structure is the same as in (\ref{espre}), upon
suitable symbolic replacements. Tracing, we get the 
same recursion relation 
as (\ref{n}), but with $n\rightarrow n+1$. The solution is
therefore
\[
b_{k}=(-1)^{k}\frac{s!~\left( s-k+\frac{n}{2}-2\right) !}{%
2^{2k}k!~(s-2k)!~\left( s+\frac{n}{2}-2\right) !}. 
\]

We have ${\cal I}_{\{\mu \},\{\nu \}}^{(s)}(x)=(-1)^{s}2^{s}
x_{1}\cdots x_{2s}/|x|^{2s}+\delta \text{-terms,}$ and the
complete conversion formula reads
\[
{\prod }_{\{\mu \},\{\nu \}}^{(s)}\left( \frac{1}{|x|^{2n-4}}\right) = \frac{%
(-1)^s2^{2s}(2s+n-3)!(s+n-4)!\left(s+{\frac{n}{2}}-2\right)!}{(n-3)! \left({%
\frac{n}{2}}-2\right)!(2s+n-4)!} {\frac{{\cal I}_{\{\mu \},\{\nu \}}^{(s)}(x)%
}{|x|^{2s+2n-4}}}. 
\]

The square of the conformal tensor ${\cal I}_{\{\mu \},\{\nu \}}^{(s)}(x)$
is equal to the identity, denoted by $\Im _{\{\mu \},\{\nu \}}^{(s)}$, in
the space of symmetric, traceless, $s$-indexed tensors. Precisely, 
\begin{equation}
{\cal I}_{\{\mu \},\{\rho \}}^{(s)}(x)~{\cal I}_{\{\rho \},\{\nu
\}}^{(s)}(x)=p_{s}^{2}~\Im _{\{\mu \},\{\nu \}}^{(s)},  \label{eqo}
\end{equation}
for some factor $p_{s}$.

We write 
\[
\Im _{\{\mu \},\{\nu \}}^{(s)}=\sum_{k=0}^{\left[ \frac{s}{2}\right]
}c_{k}\delta _{\bar\mu\bar\mu}^{k}\delta _{\bar\nu\bar\nu}^{k}\delta
_{\bar\mu\bar\nu}^{s-2k}. 
\]
The tracelessness condition shows immediately that the $c_{k}$ are
proportional to the $b_{k}$. The normalization of $\Im _{\{\mu \},\{\nu
\}}^{(s)}$ is fixed by the condition 
\[
\Im _{\{\mu \},\{\rho \}}^{(s)}~\Im _{\{\rho \},\{\nu \}}^{(s)}=\Im _{\{\mu
\},\{\nu \}}^{(s)}. 
\]
This is easy to calculate. Indeed, only the first term of $\Im _{\{\mu
\},\{\rho \}}^{(s)}$, namely $c_{0}\delta _{\bar\mu\bar\rho}^{s}$, gives a
non-vanishing result when acting on $\Im _{\{\rho \},\{\nu \}}^{(s)}$, since all
the other terms contain at least one trace. Therefore $c_{0}=1$ and the $%
c_{k}$ are precisely equal to the $b_{k}$ of (\ref{bk}).

Now we study (\ref{eqo}). Only the term $b_{0}{\cal I}%
_{\bar\mu\bar\rho}^{s}(x)$ of ${\cal I}_{\{\mu \},\{\rho \}}^{(s)}(x)$ gives
a non-vanishing contribution when acting on ${\cal I}_{\{\rho \},\{\nu
\}}^{(s)}(x)$. On the other hand, the product ${\cal I}_{\bar\mu\bar%
\rho}^{s}(x){\cal I}_{\bar\rho\bar \nu }^{s-2k}(x)$ with the $s-2k$ $\rho $%
-indices of ${\cal I}_{\bar{\rho }\bar\nu}^{s-2k}(x)$ contracted with
corresponding indices in ${\cal I}_{\bar\mu\bar\rho}^{s}(x)$ is equal to $%
\delta _{\bar\mu\bar\nu}^{s-2k}{\cal I}_{\bar\mu\bar{\rho }}^{2k}(x)$.
Further multiplication by $\delta _{\bar\rho\bar\rho}^{k}$ gives $\delta
_{\bar\mu\bar\nu}^{s-2k}\delta _{\bar\mu\bar\mu}^{k}$. In conclusion,
$p_{s}=1$.

The conformal tensor ${\cal I}_{\{\mu \},\{\nu \}}^{(s)}(x)$ provides a
universal way to normalize the two-point functions.

\section{Trace anomaly}

In this section we work out the contribution of the
``generalized'' vector field to the first central charge, called $c$,
in the trace anomaly. The
charge $c$ provides the {\it integrated} trace anomaly
at the quadratic level in the expansion around flat space.
The topological invariants and the more 
important central charge $a$ (see \cite{proceeding}
for the definition and a quick introduction), instead, 
are not visible in this formula.

The stress tensor in arbitrary dimension reads
\begin{eqnarray*}
T_{\mu\nu}&=&-{1\over 4}\varphi \overleftrightarrow{\partial_\mu}
\overleftrightarrow{\partial_\nu} \varphi+{1\over 4}
(\bar\psi\gamma_\mu \overleftrightarrow{\partial_\nu} \psi+
\bar\psi\gamma_\nu \overleftrightarrow{\partial_\mu} \psi)
+{1\over 
4(n-1)}\pi_{\mu\nu}(\varphi^2)+
F_{\mu\{\alpha\}}F_{\nu\{\alpha\}}-{1\over n}\delta_{\mu\nu}
F^2\\
&&=
-{1\over 4}{\cal J}^S_2+{1\over 2}{\cal J}^F_2-2{\cal J}^V_2,
\end{eqnarray*}
where the currents ${\cal J}^{V,F,S}$ are in the normalization of \cite{high} 
and we assume that there are $N_S$ real scalar fields
$\varphi$, $N_F$
Dirac fermions $\psi$ and $N_V$ $(n/2-1)$-forms 
$A_{\mu_1\ldots \mu_{l-1}}$,
$n=2l$.

We get
\[
\langle T_{\mu\nu}(x)\,
T_{\rho\sigma}(0)\rangle=
c_n{\left({n\over 2}\right)!\left({n\over 2}-1\right)!
\over (2\pi)^n(n+1)!}{\prod}^{(2)}_{\mu\nu,\rho\sigma}
\Box^{{n\over 2}-2}\left({1\over |x|^n}\right),
\]
where ${\prod}^{(2)}_{\mu\nu,\rho\sigma}={1\over 2}
(\pi_{\mu\rho}\pi_{\nu\sigma}+\pi_{\mu\sigma}\pi_{\nu\rho})
-{1\over n-1}\pi_{\mu\nu}\pi_{\rho\sigma}$ is the spin-2 projector
and
\[
c_n=N_S+2^{{n\over 2}-1}(n-1)N_F+{n!\over
2 \left[\left({n\over 2}-1\right)!\right]^2}N_V
\]
is the desired central charge. The result matches with the 
well-known four-dimensional expression
$c=N_S+6N_F+12N_V$.
Using
\[
\int \Theta=-\mu{\partial\over \partial\mu},~~~~~~~~~~~~~~~~
\mu{\partial\over \partial\mu}\left({1\over|x|^n}\right)=
{2\pi^{n\over 2}\over \Gamma\left({n\over 2}\right)}
\delta(x),
\]
wehre $\Theta$ denotes the trace of the stress tensor, we have also
\begin{eqnarray*}
\int \Theta&=&-{c_n\over 4}{\left({n\over 2}\right)!\over
(4\pi)^{n\over 2}(n+1)!}\int h_{\mu\nu}
{\prod}^{(2)}_{\mu\nu,\rho\sigma}\Box^{{n\over 2}-2}h_{\rho\sigma}
+{\cal O}(h^3)
\nonumber\\&=&
-{c_n\over 4}{n-2\over n-3}{\left({n\over 2}\right)!\over
(4\pi)^{n\over 2}(n+1)!}\int W
\Box^{{n\over 2}-2}W+\cdots,
\end{eqnarray*}
the expression on the right-hand side being the first term of the
expansion of the conformal invariant $\hbox{$W_{\mu\nu\rho\sigma}
\Box^{{n\over 2}-2}W^{\mu\nu\rho\sigma}+\cdots$}$
($W$ denoting the Weyl tensor - see \cite{6d} for
other details) around the flat metric
($g_{\mu\nu}=\delta_{\mu\nu}+h_{\mu\nu}$).

\section{Complete form of the higher-spin currents}

Finally, we give the complete expressions of the higher-spin currents,
including the $\delta $-terms. The derivation follows the strategy of the
previous sections, which should be familiar by now, and therefore we just
report the result.

We can write in compact notation, for a complex scalar field, 
\[
{\cal J}_{s}^{S}=\sum_{k=0}^{\left[ \frac{s}{2}\right] }a_{k}~\pi
^{k}~\left( \bar{\varphi}~\overleftrightarrow{\partial }^{(s-2k)}\varphi
\right) , 
\]
$\pi $ being the conserved projector $\partial \partial -\Box \delta $, as
usual. Here we choose $a_{0}=1$. Conservation of ${\cal J}^{(s)}$ is implicit.
Tracelessness gives the same recursion relation as in (\ref{n}). The
coefficients for the spinor and vector cases are easily obtained from 
the scalar ones
with the replacements $s\rightarrow s-1$, $n\rightarrow n+2$ and $%
s\rightarrow s-2$, $n\rightarrow n+4$, respectively. In conclusion, 
\begin{eqnarray*}
{\cal J}_{s}^{S} &=&\frac{s!~\left( s+\frac{n}{2}-2\right) !}{(2s+n-4)!}%
\sum_{k=0}^{\left[ \frac{s}{2}\right] }\frac{(-1)^{k}(2s-2k+n-4)!~}{%
k!~(s-2k)!~\left( s-k+\frac{n}{2}-2\right) !~}~\pi ^{k}~\left( \bar{\varphi}~%
\overleftrightarrow{\partial }^{(s-2k)}\varphi \right) , \\
{\cal J}_{s}^{F} &=&\frac{(s-1)!~\left( s+\frac{n}{2}-2\right) !}{(2s+n-4)!}%
\sum_{k=0}^{\left[ \frac{s-1}{2}\right] }\frac{(-1)^{k}(2s-2k+n-4)!~}{%
k!~(s-2k-1)!~\left( s-k+\frac{n}{2}-2\right) !~}~\pi ^{k}~\left( \overline{%
\psi }\gamma \overleftrightarrow{\partial }^{(s-2k-1)}\psi \right) , \\
{\cal J}_{s}^{V} &=&\frac{(s-2)!~\left( s+\frac{n}{2}-2\right) !}{(2s+n-4)!}%
\sum_{k=0}^{\left[ \frac{s}{2}-1\right] }\frac{(-1)^{k}(2s-2k+n-4)!~}{%
k!~(s-2k-2)!~\left( s-k+\frac{n}{2}-2\right) !~}~\pi ^{k}~\left( F_{\{\alpha\}
}^{+}\overleftrightarrow{\partial }^{(s-2k-2)}F_{\{\alpha\} }^{-}\right) ,
\end{eqnarray*}
in the normalization of ref. \cite{high}. We have checked agreement with the 
explicit formulas of \cite{high}. In \cite{high} and the previous sections
${\cal J}^S_s$ was normalized with an additional factor 2.

\vskip .5truecm

{\bf Acknowledgements}

I am grateful to M. Porrati and A. Zaffaroni for discussions.

\section{Appendix: the $(n/2-1)$-form}

The propagator of the ``generalized'' conformal vector field
of section 5 is
\[
\langle A_{\{\alpha\}}(x)\, A_{\{\beta\}}(0) \rangle=
{\Gamma\left({n\over 2}\right)\over 2(n-2)\pi^{n\over 2}}
{{\cal A}_{\{\alpha\},\{\beta\}}\over |x|^{n-2}}.
\]
Here ${\cal A}_{\{\alpha\},\{\beta\}}$ is the identity in the space
of antisymmetric tensors with $l-1$ indices in $2l$ dimensions,
${\cal A}_{\{\alpha\},\{\beta\}}={1\over (l-1)!}
\delta^{\{\alpha\}}_{\{\beta\}}$ and $\delta^{\{\alpha\}}_{\{\beta\}}$
is the determinant of the matrix having Kronecher tensors
as entries,
\[
{\cal A}_{\{\alpha\},\{\beta\}}={1\over (l-1)!(l+1)!}
\varepsilon_{\{\alpha\}\gamma_1\ldots \gamma_{l+1}}
\varepsilon_{\{\beta\}\gamma_1\ldots \gamma_{l+1}}. 
\]
Note that the formula for the trace:
\[
{\cal A}_{\{\alpha\},\{\alpha\}}={(2l)!\over (l-1)!(l+1)!}.
\]
It is possible to define a dual field-strength,
\[
\tilde F_{\mu_1\ldots \mu_l}={1\over l!}
\varepsilon_{\mu_1\ldots \mu_l\nu_1\ldots \nu_l}F^{\nu_1\ldots \nu_l},
\]
as well as self-dual and anti-self-dual tensors
\[
F^{\pm}_{\mu_1\ldots \mu_l}={1\over 2}
(F_{\mu_1\ldots\mu_l}\pm \tilde F_{\mu_1\ldots \mu_l}), 
\]
so that the stress tensor reads
\[
T_{\mu\nu}=F^+_{\mu\{\alpha\}}F^-_{\nu\{\alpha\}}+
F^-_{\mu\{\alpha\}}F^+_{\nu\{\alpha\}}.
\]
Observe that only in dimension multiple of 4,
the two terms of this sum are equal.
Indeed, we have
\begin{eqnarray}
M^{\pm}_{\mu_1\ldots\mu_l}&=&\pm{1\over l!}\varepsilon_{
\mu_1\ldots \mu_l\nu_1\ldots \nu_l}M^{\pm}_{\nu_1\ldots\nu_l},~~~~~~~~~~~~
~~~~~~~~{\rm for} \,\,\, l{\rm =even},
\label{unouno}
\\
M^{\pm}_{\mu_1\ldots\mu_l}&=&\pm{1\over l!}\varepsilon_{
\mu_1\ldots \mu_l\nu_1\ldots \nu_l}M^{\mp}_{\nu_1\ldots\nu_l},~~~~~~~~~~~~
~~~~~~~~{\rm for} \,\,\, l{\rm =odd},
\label{duedue}
\end{eqnarray}
where $M^{\pm}$ is any self-dual/anti-self-dual 
tensor in the sense specified above.

The field equations $\partial^\mu F_{\mu\alpha_1\ldots\alpha_{l-1}}=0$
and Bianchi identities
$\partial^\mu \tilde F_{\mu\alpha_1\ldots\alpha_{l-1}}=0$
can be written in the form
$\partial^\mu F^{\pm}_{\mu\alpha_1\ldots\alpha_{l-1}}=0$.
For $l$=even, we can write the stress tensor as
\[
T_{\mu\nu}=2F^+_{\mu\{\alpha\}}F^-_{\nu\{\alpha\}},
\]
without care about the positions of the indices $\mu$ and $\nu$.
The same can be said of the currents with arbitrary spin.
The proof of this statement follows from the equality
\begin{equation}
M^+_{\mu\{\alpha\}}N^-_{\nu\{\alpha\}}=M^+_{\nu\{\alpha\}}N^-_{\mu\{\alpha\}},
~~~~~~~~~~~~~~~~~~~~(l{\rm =even}),
\label{cross}
\end{equation}
which can be shown using (\ref{unouno}) for $l$=even, with
the bonus
\begin{equation}
M^+_{\mu_1\ldots \mu_l}N^-_{\mu_1\ldots \nu_l}=0
\label{trace}
\end{equation}
Here $M$ and $N$ are generic self-dual and anti-self-dual tensors,
respectively.

In $l$=even conservation of the stress-tensor is obvious
and tracelessness follows from (\ref{trace}).
This means that the current-formalism of four dimensions generalizes straightforwardly
to arbitrary dimension multiple of four.

If $l$ is even the instanton equation
\begin{equation}
F^+_{\mu_1\ldots \mu_l}=0
\label{inst}
\end{equation}
is meaningful. If $l$ is odd, instead, (\ref{duedue}) implies
that self-dual and anti-self-dual tensors are not independent
of each other, so
that if (\ref{inst}) holds, then
$F^-_{\mu_1\ldots \mu_l}$ also vanisehs and there
are only trivial solutions.
It is possible, nevertheless, to have a nontrivial notion of instanton
in $l$=odd by considering complex fields. 

So, let us assume that for $l$=odd the field
is complex. In real notation we have two fields
$A_{\mu_1\ldots \mu_{l-1}}^i$, $i=1,2$. The field strength
$F^i_{\mu_1\ldots \mu_l}$ is defined as above, but now
\[
F^{i\pm}_{\mu_1\ldots \mu_l}={1\over 2}
\left(
F^i_{\mu_1\ldots \mu_l}\pm
{1\over l!}\varepsilon^{ij}\varepsilon_{
\mu_1\ldots \mu_l\nu_1\ldots \nu_l}F^j_{\nu_1\ldots \nu_l}
\right).
\]
Now we have a good version of (\ref{unouno}),
\[
M^{\pm i}_{\mu_1\ldots\mu_l}=\pm{1\over l!}
\varepsilon^{ij}\varepsilon_{
\mu_1\ldots \mu_l\nu_1\ldots \nu_l}M^{\pm j}_{\nu_1\ldots\nu_l}.
\]
The stress tensor reads
\[
T_{\mu\nu}=2F^{+i}_{\mu\{\alpha\}}F^{-i}_{\nu\{\alpha\}}
\]
and everything else generalizes immediately, in particular
formulas (\ref{cross}) and (\ref{trace}).
The instanton equations are non-trivial.

All this is natural, because
in two dimensions our ``generalized vector field'' reduces
to the free (complex) scalar $\varphi$. 
Self-duality (anti-self-duality) of the ``field strength''
$F_{\mu}^{+i}=0$ ($F_{\mu}^{-i}=0$) is the holomorphicity
(anti-holomorphicity) condition on $\varphi$. These observations
might be useful to estabilish correspondences
between instantons in different dimensions.

The instanton conditions (\ref{inst}) have non-trivial solutions on
non-trivial manifolds, upon covariantization
of the theory, with
topological invariant
\[
\int {\rm d}^nx\, F_{\mu_1\ldots\mu_l} \tilde F^{\mu_1\ldots \mu_l}.
\]

\end{document}